\documentclass[article]{aa}
\usepackage{psfig,latexsym,longtable,lscape,supertabular}

\def\e20{$\times 10^{20}$}
\def\ergsec{erg s$^{-1}$}
\def\ergcmsec{erg cm$^{-2}$ s$^{-1}$}

\def\today{\number\day -\number\month -\number\year}
\def\chandra{{\it Chandra}}

\def\msol{{\rm M}_{\sun}}
\def\hi{H\,{\sc i}\,}
\def\nii{N\,{\sc i}\,{\sc i}\,}
\def\hii{H\,{\sc i}\,{\sc i}\,}
\def\sii{S\,{\sc i}\,{\sc i}\,}

\begin{document}

\pagenumbering{arabic}
\title{Stephan's Quintet: The X-ray Anatomy of a Multiple Galaxy Collision}

\author{Ginevra Trinchieri\inst{1}, Jack Sulentic\inst{2},
Dieter Breitschwerdt\inst{3}
and
Wolfgang Pietsch\inst{3} }
\institute{
INAF-Osservatorio Astronomico di Brera, via Brera 28, 20121
 Milano Italy
 \and
Physics \& Astronomy, University of Alabama, USA
\and
 Max-Planck-Institut f\"ur extraterrestrische Physik,
              Giessenbachstra\ss e, D-85740 Garching
              Germany
}
   \offprints{G.~Trinchieri}
   \mail{ginevra@brera.mi.astro.it}

   \date{Draft: \today}%Received date; accepted date}
\authorrunning{Trinchieri et al.}
\titlerunning{Shock in SQ}

\abstract{Chandra observations of the compact galaxy group known as
Stephan's Quintet (SQ) are presented. The major morphological features
that were discovered with the ROSAT HRI are now imaged with higher
resolution and S/N.  The large scale shock (1$\farcm$5, $\sim$40kpc if at
85 Mpc) is resolved into a narrow NS feature embedded in more extended
diffuse emission (D$\ge3'$).  The  NS structure is  somewhat clumpy,
more sharply bounded on the W side and prominent only in the soft band
(energies below $\sim2$ keV). Its observational properties are best
explained as a shock produced by a high velocity encounter between
NGC~7318b, a ``new intruder'',  and the intergalactic medium in SQ.  The
shock conditions near the high speed intruder suggest that a bow shock
is propagating into a pre-existing \hi\ cloud and heating the gas to a
temperature of $~ 0.5$ keV.  The low temperature in the shock is a
problem unless we postulate an oblique shock. 
One member, NGC~7319, hosts a Seyfert 2
nucleus,  with an intrinsic luminosity L$_X$= $10^{43}$ \ergsec,
embedded in a region of more diffuse emission with 10$''$ radius
extent.  The nuclear spectrum can be modeled with a strongly absorbed
power-law typical of this class of sources.  Several additional compact
sources are detected including three in foreground NGC~7320. Some of
these sources are very luminous and could be related to the
ultraluminous X-ray sources found in nearby galaxies.
 }

\maketitle

\section{Introduction}

Stephan's Quintet (HCG92, Hickson 1982; hereafter SQ) is the most studied
example of the compact group phenomenon. It is composed of six galaxies
(Sulentic et al. 2001; S01 hereafter) including a core of three
(NGC~7317, NGC~7318a and NGC~7319) with essentially zero velocity
dispersion and an unrelated foreground object (NGC~7320). 
Multiwavelength observations of SQ give strong evidence of multiple
episodes of past and recent (current) interaction, most likely caused by
acquisition of new members/passage of intruders 
from the associated larger scale galaxy population near SQ. 
Both NGC~7320c and NGC~7319 show
spiral morphology without detectable \hi\, while the other two core
members are ellipticals.  The last and presumably
ongoing event involves the collision of
the gas-rich spiral NGC~7318b with the debris field produced by past
interactions.  About half of that galaxy's  ISM has been
stripped/shocked. SQ is certainly the best local example of a compact
group caught {\it in flagrante delicto} with multiple manifestations of
interaction events.  It is this kind of event that presumably accounts
for general compact group characteristics including \hi\ deficiency,
quenched star formation and an excess of early-type members (e.g.
Hickson \& Rood 1988; Sulentic \& de Mello Rabaca 1993;
Verdes-Montenegro et al. 2001; Vilchez \& Iglesias-Paramo 1998).  SQ is
thus an ideal laboratory for studying interactions {\it per se}, as
well as, an excellent local analog to the processes thought to be much
more common at high redshift.

We present CHANDRA observations of SQ  where we confirm the complex
nature of the X-ray emission already reported (Pietsch et al. 1997, S01). 
Most, if not all, of the basic types of
X-ray emission observed from extragalactic sources are found in SQ
(AGN, shocks, jets, ``normal'' galactic emission, diffuse emission), and
can be studied in this context.

%--------------------------Figure 1 large --------------------------------
\begin{figure*}
\resizebox{18cm}{!}{
\psfig{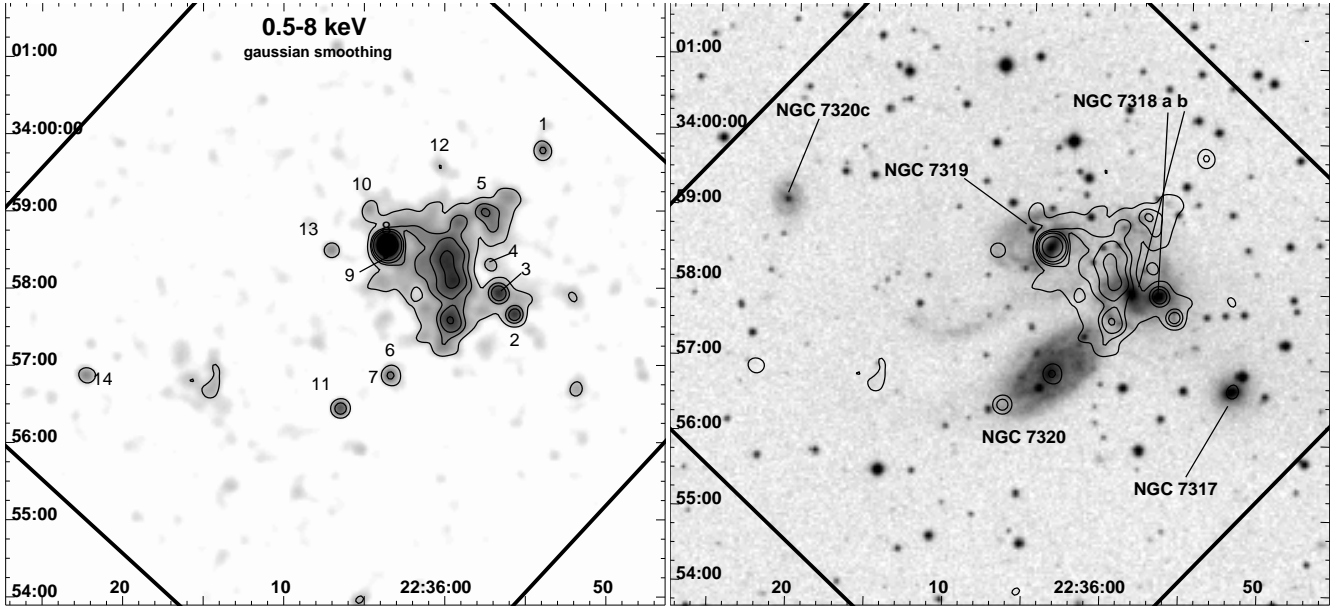}
}
\caption{Left: Full field Chandra image (ACIS-S: back-illuminated chip
S3 only) of SQ in the 0.5-8 keV energy range with X-ray
contours and source numbers (see \S~\ref{sourcesection}) superimposed.
The map is binned with $1''\times 1''$ pixels
and smoothed with a bidimensional Gaussian filter ($\sigma=4''$).
Right: X-ray contours superimposed on the blue DSS2 image of the
field.  Galaxies are identified, and the ACIS-S field of view is indicated.}
\label{large}
\end{figure*}
%-------------------------end Figure --------------------------------

%--------------------------Figure xopt -----------------------------------
\begin{figure}
\psfig{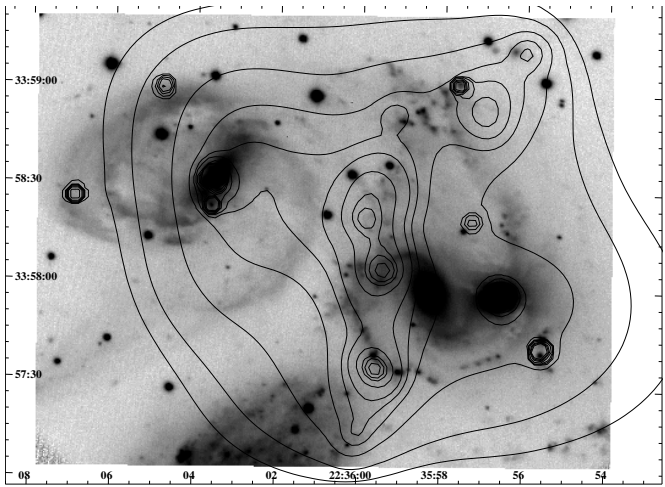}
\caption{A zoomed X-ray contour map (0.5-3 keV) of the main concentration of
X-ray photons. Contours are superimposed on a CFHT
B-band image of the field.  The X-ray data are
smoothed with an adaptive filter and 2.5$\sigma$ as the lowest
significance of the signal within the kernel. 
}
\label{xopt}
\end{figure}
%-------------------------end Figure --------------------------------

\section{Basic results of the \chandra\ data analysis}

A 19.7 ks observation of SQ was obtained with \chandra\ in July 2000
using the back-illuminated CCD (ACIS-S in imaging configuration).  The
data were reprocessed using the new calibration files as described in
the CIAO documentation. Following the ``CIAO Science Threads" at the
CXC home page\footnote{\tt
http://asc.harvard.edu/ciao/threads/threads.html}, we verified that
the  data obtained with standard processing had been properly cleaned
(e.g. for high background events). We produced several images with $1''
\times 1''$ pixel resolution, covering different portions of the field
and in different energy bands.   We applied a Gaussian filter or an
adaptive smoothing filter using the {\tt csmooth} routine in the CIAO
package ({\tt FFT} method, and using 2.5$\sigma$ as the minimal
significance of the signal within the kernel).  The left panel of
Figure~\ref{large} presents an X-ray image of the entire
back-illuminated CCD chip (S3) with X-ray contours superimposed.  The
right panel shows X-ray contours superposed on an optical B-band image
from the DSS2.  Considerable extended emission is concentrated within
the core of the compact group.  Additional sources are associated with
the galaxies (members of SQ and the foreground galaxy NGC~7320) or, in
some cases, may be unrelated background sources (see
Table~\ref{sourcelist}).

Fig.~\ref{xopt} presents a closer look at the X-ray emission most
unambiguously associated with SQ. The soft (0.5-3.0 keV) X-ray contours
are shown overlayed on an average of CFHT B-band images kindly provided
by C. Mendes de Oliveira (see Plana et al. 1999; Mendes de Oliveira et
al. 2001; S01 for discussion of the images).  The complex X-ray
emission that was detected in previous ROSAT observations (Pietsch et
al.  1997; S01) is clearly resolved into two main components almost
certainly associated with SQ:  1) complex clumpy and extended emission
centered on a radio continuum/optical emission-line emitting shock zone
and 2) emission from the Seyfert nucleus in NGC~7319. 
Additional sources detected do not always  have obvious optical
counterparts, although they appear to be associated with the galaxies
in several cases (the central region of
NGC~7318a, and associations in  Table~\ref{sourcelist}).

\begin{figure*}
\resizebox{18cm}{!}
{
\psfig{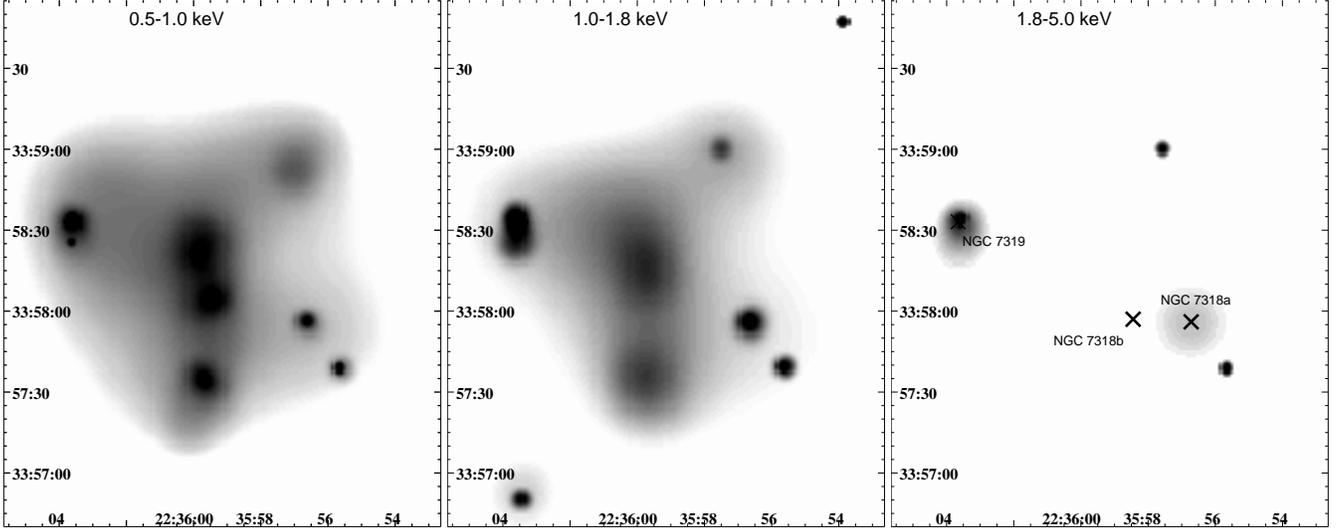}
}
\caption{
Smoothed images of the ACIS-S data in SQ, in different energy bands.
An adaptive filtering technique (FFT convolution method) is applied
to the data.  {\bf Crosses in the rightmost panel indicate the galaxy's
centers.}
}
\label{manye}
\end{figure*}

Fig.~\ref{manye} shows \chandra\ images in different energy ranges.
Comparison of the images shows that the extended X-ray emission is
found only below $\sim 2$ keV, and more compact sources coincident with the
Seyfert 2 nucleus of NGC~7319 and an unresolved source SW of the
NGC~7318a nucleus and apparently coincident with one of the new intruder
emission regions (\# 14 in Fig. 7 of S01) are prominent at higher
energies.

%--------------------------Figure 3 manye-----------------------------------
%-------------------------end Figure --------------------------------

\subsection{Distribution of Extended X-ray Emission}
\label{profiles}

\begin{figure}
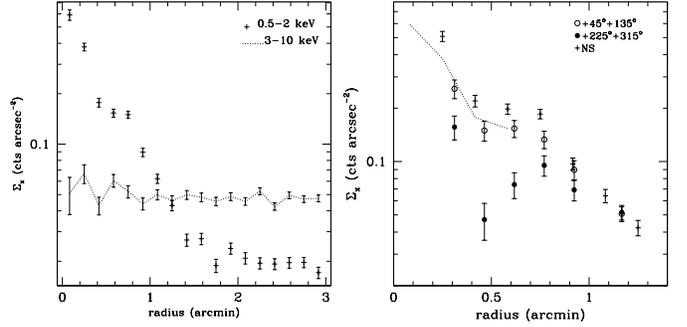

\resizebox{9cm}{!}{
\psfig{file=ms3179.f4a,width=9cm,clip=}
\psfig{file=ms3179.f4b,width=9cm,clip=}
}
\caption{LEFT: Azimuthally averaged radial profiles of the total emission
in  the 0.5-2.0 and 3.0-10.0keV energy ranges.
RIGHT: Radial profile of the total emission in different
azimuthal quadrants, as indicated
(0.5-2 keV band).  The $\pm 45\degr$ azimuthal profiles along
the N and S directions are very similar and have been averaged together.
Only the radial range $0\farcm2-1\farcm5$ is shown because the profiles are 
the same beyond that range. The 360$\degr$
azimuthal average at the center is also shown as a dotted line.
}
\label{angprof}
\end{figure}

Diffuse emission in SQ extends from NE of 
the Seyfert 2 nucleus of NGC~7319 to SW of 
the nucleus of NGC~7318a.  Elongated clumpy 
NS structure lies near the center of more 
diffuse emission. We derived azimuthally averaged 
radial profiles of the extended emission
in several energy bands and in several ranges 
of position angle. The goal of the radial profile
analysis was to determine the extent of diffuse 
emission for comparison with images at other 
wavelengths. We centered the radial profiles at 
RA=22$^h$ 35$^m$ 59.7$^s$, $\delta$=33$^\circ$ 58$^\prime$
14.1$^{\prime\prime}$ and excluded all discrete 
sources (\S~\ref{sourcesection}). Figure~\ref{angprof} 
shows the azimuthally averaged 
raw count profiles in two broad energy bands. 
Extended emission is only detected 
below $\sim 1.8$ keV and out to a radius of r$\sim$2$^\prime$.

%------------------------- Figure angprof -------------------
%------------------------- Figure angprof -------------------

%------------------------- Figure cut-------------------
\begin{figure}
\psfig{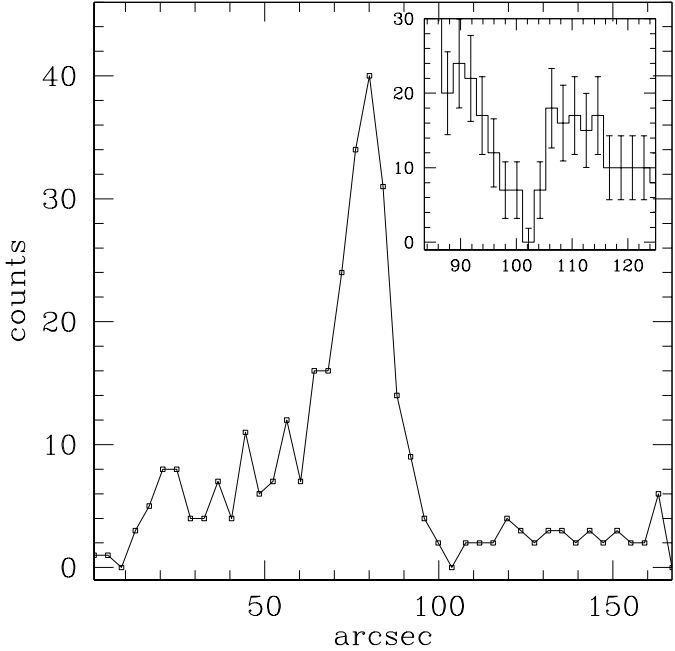}
\caption{Total counts in a EW oriented strip
centered at 22$^h$35$^m$59.36$^s$,
+33$^\circ$58$^\prime$07.2$^{\prime\prime}$.
The horizontal scale covers $170''$,
and each point plotted represents the (0.5-2 keV band) counts in bins of size
$\sim 4'' \times 12''$.
The inset shows an
enlargement  of the region around r=$100''$ corresponding to the  minimum intensity,
and relative error bars, which was obtained in a much larger area, so that each step
represents the counts in $\sim 2'' \times 62\farcs5$ regions.
Counts associated with detected  sources are not included.
}
\label{cut}
\end{figure}
%------------------------- Figure cut-------------------

Profiles derived for different azimuthal quadrants 
indicate that the extent of the emission is similar 
in all directions but that the intensity distribution 
is not. Detailed comparison of azimuthal averages is 
not straightforward because the extended emission is 
complex. The NS distribution is dominated by the shock. 
The E profile is similar to the NS 
distribution although less intense. It shows a smoother 
decline and a higher intensity level than towards the W. 
A larger and more abrupt drop in intensity is seen 
towards the W followed by an apparent rebound (Fig~\ref{angprof}). 
The E-W difference is also evident in 
the EW cut taken across the middle intensity peak of the 
NS feature (Fig.~\ref{cut}). This plot indicates a sharp decrease to a 
very low count level in a narrow and elongated region 
($\sim8'' \times 1'$) along the W edge of the NS 
feature. The emission increases again W of this minimum 
and extends outwards for an additional $50''-70''$.
The low intensity strip is not aligned with the pixels 
on the CCD chip which leads us to conclude that it is 
not an instrumental artifact. 

An additional weak extended X-ray component is seen near
RA=22$^h$36$^m$15$^s$,
$\delta$=+33$^\circ$56$^\prime$30$^{\prime\prime}$.  This 15$''$ radius
region ($\sim2-3\sigma$ source) coincides with structure in the older
tidal tails (Arp \& Lorre 1976; S01).  However, since several optical
condensations are found in this same region we cannot rule out a line
of sight coincidence with background sources. We will explore this
feature in more detail with recently obtained XMM-Newton data
(Trinchieri et al. in prep).

\subsection{Discrete X-ray Sources}
\label{sourcesection}

A considerable number of discrete sources is found in and near SQ.
Some are resolved and other are not, while several show  associated
(i.e. concentric) extended emission. Table~\ref{sourcelist} lists
discrete sources with positions, count statistics, fluxes, X-ray
luminosities and possible optical counterparts (see Fig.~\ref{large}).
The source list is derived from the {\tt wavdetect} and {\tt
celldetect} programs that gave virtually identical results.  The
background is derived locally around each source and is negligible for
sources outside the region where complex emission is detected.  
Count rates in the total energy band are
converted into unabsorbed fluxes and luminosities assuming a constant
conversion rate of 1 c $s^{-1}$ = 9$\times 10^{-12}$ erg cm$^{-2}$ s$^{-1}$
(corresponding to either a power law with $\Gamma$=1.7, or a
bremsstrahlung with kT=10 keV, and line-of-sight absorption), and a
distance of 85.6 Mpc for SQ sources, and 12 Mpc for
NGC~7320 (see Table~\ref{sourcelist}).  Only fluxes are quoted for
sources outside or not unambiguously associated with SQ.

\begin{table*}
\caption[]{Positions, net counts in the broad energy band (0.3-10 keV), 
fluxes and luminosities (0.3-10 keV) in c.g.s. units 
of the sources detected in SQ.
Net counts are derived in a $1''$ radius circle above the local
background (except for source \# 14, for which r=3$''$, and 
converted into fluxes assuming a conversion factor 4.6$\times 10^{-16}$,
corresponding to a Bremsstrahlung model of 10 keV
except for the nuclear region in NGC 7319, where the best fit from the
spectral analysis is used.
}
\begin{tabular}{lllrrlll}
\hline
\hline

 CXC name& Sou & RA / Dec & Total & Net & flux & luminosity& Notes \\
&\multicolumn{1}{c}{\#} & \multicolumn{1}{c}{(J2000)} & 
\multicolumn{2}{c}{counts} & \multicolumn{2}{c}{(0.3-10 keV)}\\
CXOU J223553.9+335946&1  & 22:35:53.94 &  24 &  23.8 $\pm$   
4.9 & 1.1$\times
   10^{-14}$ & & Background? \\
   &&   33:59:46.91    &&&&& no optical counterpart\\
CXOU J223555.6+335738 & 2 & 22:35:55.69 &  58 & 57.6  $\pm$   7.6  & 2.7$\times
   10^{-14}$    & & Associated with  \\
   &&   33:57:38.82 &&&&&  NGC 7318b?\\
CXOU J223556.6+335756 & 3 & 22:35:56.68 &  27 &  24.4  $\pm$ 5.1 &1.1$\times
   10^{-14}$& 9.7$\times 10^{39}$ & NGC 7318a 1$''$ radius \\
&    &  33:57:56.04 & 72&  64.4 $\pm$ 8.5 & 3.0$\times10^{-14}$& 2.7$\times 10^{40}$&
     NGC 7318a 5$''$ radius \\
CXOU J223557.3+335818 & 4 & 22:35:57.36 &  8  &  7.4 $\pm$    2.8 &3.4$\times
   10^{-15}$&     & Near NGC7318b \\
   &     & 33:58:18.76 &&&&& emission reg \\
CXOU J223557.6+335859 &   5  &22:35:57.61 &  26 &  25.4 $\pm$    5.0 &  
     1.2$\times 10^{-14}$&  & Near NGC7318b \\
     &     & 33:58:59.72  &&&&& emission reg\\
CXOU J223603.4+335653 &   6  & 22:36:03.40 & 13  & 12.5 $\pm$    3.6 & 
      5.8$\times 10^{-15}$& 9.9$\times 10^{37}$    & NGC 7320 nucleus?  \\
	&     & 33:56:53.74 &&&&& \\
CXOU J223603.4+335650 &   7  & 22:36:03.46 & 15 &  14.6 $\pm$    4.1 & 
      6.7$\times 10^{-15}$& 1.2$\times 10^{38}$  & NGC 7320 \\
	&     & 33:56:50.40 &&&&&  S of nucleus\\
CXOU J223603.6+335833 &   8  & 22:36:03.60 &  706&  699 $\pm$   26.6 & 
      \llap{(}1.4$\times 10^{-11a}$\rlap{)}&  \llap{(}1.1$\times
	10^{43a}$\rlap{)}& NGC 7319 - nucleus   \\
     & & 33:58:33.12 & & &  \llap{(}2.6$\times 10^{-14b}$\rlap{)} & \llap{(}2.3$\times 10^{40b}$\rlap{)} & NGC
      7319 - extended \\
CXOU J223603.6+335825 & 9  &22:36:03.66 &  36 &  34.7 $\pm$    6.0 &  
      1.6$\times 10^{-14}$& 1.5$\times 10^{40}$ & NGC 7319 \\
	&     & 33:58:25.01 &&&&&  S of nucleus \\
CXOU J223604.8+335901&10  & 22:36:04.82 &  13 &  12.6 $\pm$    3.6 &   
      5.8$\times 10^{-15}$&5.1$\times 10^{39}$  & NGC 7319  \\
	&     & 33:59:01.19&&&&&  \\
CXOU J223606.5+335625 &11  &22:36:06.50 &   25 &  24.4 $\pm$    5.0 & 
     1.1$\times 10^{-14}$ & 1.9$\times 10^{38}$ & NGC 7320 \\
     &     & 33:56:25.83 &&&&& SE edge \\
CXOU J223606.7+340106 &  12  &22:36:06.70 & 8  &  7.8 $\pm$    2.8 & 3.6$\times
  10^{-15}$ & & Background? \\
  &     & 34:01:06.40  &&&&& no optical counterpart \\
CXOU J223607.0+335829 &  13  & 22:36:07.05 &  18 &  17.5 $\pm$    4.2 &  
    8.1$\times 10^{-15}$&7.2$\times 10^{39}$  & NGC 7319 \\
    &     & 33:58:29.52 &&&&& E source  \\
CXOU J223622.2+335651 &  14  & 22:36:22.29 & 19 &  13.3 $\pm$    4.7 & 
      6.1$\times 10^{-15}$&  & Background? \\
	&     & 33:56:51.70 &&&&& no optical counterpart\\

\hline \hline
\end{tabular}
\label{sourcelist}

NOTES: \\
$^a$ The flux is derived from the spectral analysis and refers to
the nuclear component only (see \S~\ref{energydist}).  Counts refer to the total
emission within 10$''$ radius. \\
$^b$ The flux refers to the extended emission within a 1$0''$ radius
around the nuclear source, and is derived from the spectral analysis
(see \S \ref{energydist}).
\end{table*}

\subsection{X-ray Spectra for the Strongest Components}
\label{energydist}

Given the statistical significance of the \chandra\ data, we  
confine  our spectral analysis to the extended emission and the
central regions of NGC~7319. We used the {\tt psextract} script in CIAO
to create appropriate spectral matrixes (eg. {\em arf} and {\em rmf}
files) for the analysis, which was done using XSPEC. While the script
is not appropriate for extended emission, the regions we have selected
are small and, consequently,  the error introduced by the point-source
assumption is negligible (see also discussion in Trinchieri \&
Goudfrooij  2002; Hicks et al. 2002).  The degradation of the ACIS
Quantum efficiency, recently discussed (\footnote{\tt
http://asc.harvard.edu/cal/Acis/Cal\_prods/qeDeg}) was considered.
However, our data are not affected significantly because the observation
was done at early stages in the mission.

\subsubsection{The NS Extended Feature}
\label{NS-feature}

%------------------------- Figure shockspec  -------------------
\begin{figure}
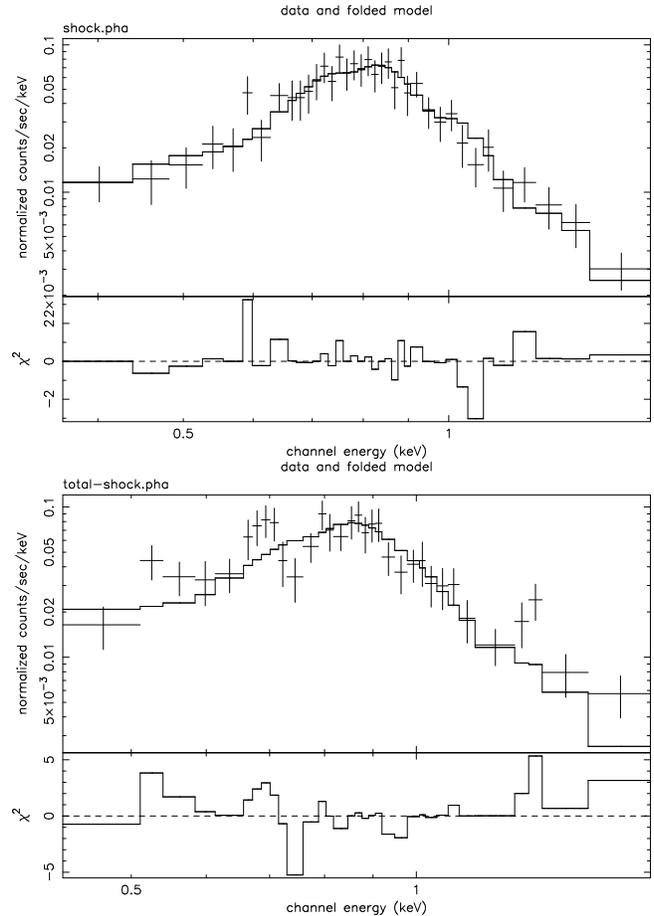

\psfig{file=ms3179.f6a,width=8.5cm,angle=-90,clip=}
\psfig{file=ms3179.f6b,width=8.5cm,angle=-90,clip=}
\caption{Spectral distribution and plots of the contributions to 
$\chi^2$ in the extended
NS source (TOP) and in the residual surrounding emission (BOTTOM).
Error bars on the y-axis indicate the statistical
uncertainty, on the X-axis the width of the energy bin.  
{\bf The model is in both cases a MEKAL with galactic line-of-sight
absorption, low ($\sim$20 \% cosmic) abundances, and  kT$\sim$0.5 keV
(TOP panel) and kT$\sim$0.6 keV (BOTTOM panel); see text.  }
}
\label{shockspec}
\end{figure}
%------------------------- Figure shock  -------------------

We identified a region along the NS feature that includes the three
brightest condensations. The background was obtained from adjacent
regions.  The data was binned to yield a  S/N $\geq$ 3 per bin after
background subtraction. This feature was only detected in the 0.35-1.75
keV energy range.  We adopted a plasma code (MEKAL model in XSPEC) to
describe the data which gives best fit parameters kT$\sim$0.53 keV
(0.47-0.58 at 90\% confidence for one interesting parameter), low
abundance (0.12-0.30) and absorption consistent with the galactic line
of sight value  of N$_H$ $\sim 8\times 10^{20}$ cm$^{-2}$ (see
Fig.~\ref{shockspec}).  Assuming the best fit model parameters, we
derive an unabsorbed f$_X (0.1-2 keV) \sim 1.7 \times 10^{-13}$ \ergcmsec,
and L$_X \sim 1.5 \times 10^{41}$ \ergsec.  Some residuals might be
present possibly even coincident with oxygen lines that are not
accounted for by the assumed single temperature model. They might be
indicative of more complex gas properties (see Fig.~\ref{shockspec}).
We also note that the derived best fit parameters are far from unique:
a two temperature component spectrum, with a 0.5 keV MEKAL plasma and
solar abundances plus $\sim 5$ keV bremsstrahlung component yield an
equally good fit to the NS feature.  Moreover, a crude examination of
the X-ray colors along the NS feature indicates that the three clumps
might have different energy distributions, suggesting spectral
variations even within this small region.

The lower surface brightness emission around the NS feature (i.e.  what
we used as background)  was also modeled (Fig.~\ref{shockspec}).  We
selected a circle of $1\farcm5$ radius, excluding discrete sources
(e.g. NGC~7318a and NGC~7319) and the NS feature.  Assuming a thermal
plasma spectrum with low abundances we  find kT=0.61 keV (0.56-0.65).
The total unabsorbed flux of this component is  f$_X
(0.1-2 keV) \sim 2.8 \times 10^{-13}$ \ergcmsec.  

We point out that our
interpretation of the NS feature and surrounding area in terms of a
recent shock would likely imply different spectral characteristics
(namely a multi-temperature non-equilibrium spectrum,
cf.\ Breitschwerdt \& Schmutzler 1999) than those derived above under
the assumptions of equilibrium conditions.
However, the relatively small number of photons in these
components does not allow us to consider more sophisticated models,
that might be possible with the higher statistics XMM-Newton
data.

%------------------------- Figure compspec------------------
\begin{figure}
\psfig{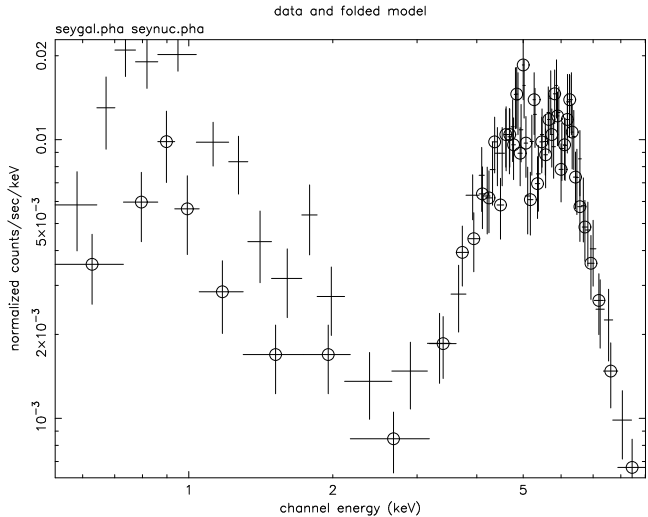}
\caption{Comparison of the energy distribution for resolved and unresolved
photons from NGC~7319. Error bars on the y-axis indicate the statistical
uncertainty, on the X-axis the width of the energy bin.  
The larger aperture ($10''$ radius, plus symbol)
indicates a significant increase in photons with energies below
2 keV relative to the smaller ($1''$ radius, circle symbols) aperture.
}
\label{compspec}
\end{figure}
%------------------------- Figure compspec------------------

%------------------------- Figure seyspec-------------------
\begin{figure}
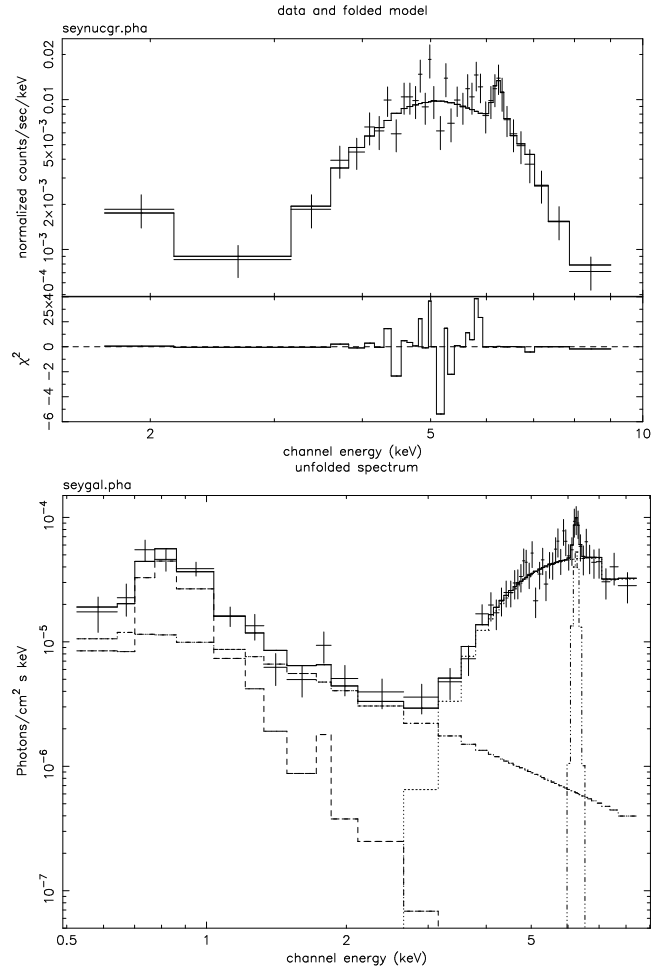

\psfig{file=ms3179.f8a,width=8.5cm,angle=-90,clip=}
\psfig{file=ms3179.f8b,width=8.5cm,angle=-90,clip=}
\caption{Energy distribution of the counts from the
the entire $10''$ radius region centered on NGC~7319 (and  $\chi^2$
contributions) are shown in the top panel.
The unfolded spectrum and the three components used in the model {\bf 
(MEKAL with kT$\sim$0.65 keV; unabsorbed power law + absorbed power
law with N$_H$ (intrinsic)$\sim$4$\times 10^{23}$ cm$^{-2}$ and
$\Gamma$$\sim$1.7; redshifted 6.4 keV Gaussian line; see text)}
are shown in the bottom panel.  }
\label{seyspec}
\end{figure}
%------------------------- Figure seyspec-------------------

\subsubsection {The nuclear region of NGC~7319}

We examined the spectral properties of both the central point source
($\sim 1''$ radius circle) and a larger region (r=10$''$) inclusive of
it. The
nuclear X-ray spectrum is complex and requires multiple components. 
Comparison with the spectral distribution for
the larger region indicates an increase in a soft contribution which
may be related to a  ``galactic" component (see Fig.~\ref{compspec}).

Since the optical spectrum indicates a Seyfert 2 nucleus, we assume a
combination of absorbed and unabsorbed power law components plus a
narrow 6.4 keV emission line (see e.g. Della Ceca et al.  1999; Moran
et al. 2001; and references therein).  The power law index should be
the same for both components with the unabsorbed component representing
$\sim 1-10$\% of the absorbed one.  Moran et al. report two different
power-law indices for an average X-ray spectrum involving 29 sources
although a single value is consistent with their data.  Given the
limited statistical significance of our data, we model the nuclear
source using a single $\Gamma$ for both power-law components (free)
plus a Gaussian line at 6.4 keV (at the redshift of NGC~7319).

We applied the absorbed+unabsorbed power-law model to both the nuclear
and larger regions. Data above $\sim 1.5-2$ keV are well described by
such a model but excess residual emission is found at lower energies
and becomes  stronger in the spectrum of the larger region.  This
motivated us to exclude data below $\sim 2$ keV from the fit.  Adopting
such a  ``nuclear" model yields best fit $\Gamma$ $\sim$ 1.7, N$_H
(intrinsic) $= 4 $\times 10^{23}$ cm$^{-2}$. The FeK line is fit with
E=6.4 keV and EW $\sim110$ eV.  This model gives a good $\chi^2_\nu \sim 1$
for 29 degrees of freedom (dof).  Fig.~\ref{seyspec} shows the spectral
data and  the contributions to $\chi^2$ in individual energy bins
relative to the combined plasma + unabsorbed and absorbed power-law
model with Gaussian line.  A bump is suspected at E $\sim 5.9$ keV.
Although introducing this additional component might improve the fit
(the best fit $\chi^2$ value reduces by 8 for 3 additional parameters;
f-test probability=99.99\%), there are no known thermal emission lines
in this energy range (see Wilms et al.  2001 where a similar feature is
suspected in a Seyfert 1 source).  The larger aperture data require an
additional component at soft energies that can be modeled with either a
MEKAL or a $raymond$ plasma code (Raymond \& Smith 1997) with kT = 0.65
keV (MEKAL) or 0.75 ($raymond$).

The data quality does not allow us to derive meaningful confidence
contours, nor does it guarantee that the model is unique. We can
however describe the spectral characteristics of the emission from
NGC~7319 as due to the superposition of a strong and heavily absorbed
nuclear source embedded in more diffuse softer emission.  The
parameters for the nucleus are consistent with those of a typical
Seyfert 2.  The intrinsic fluxes derived from the best fit models are:
f$_X \sim 4.7\times 10^{-14}$ \ergcmsec\ (0.1-2 keV) and  4$\times
10^{-14}$ \ergcmsec\ (0.3-10 keV) for the extended soft emission; f$_X
\sim  1.4 \times 10^{-11}$ \ergcmsec\ in the entire 0.3-10 keV band
and 8.2$\times 10^{-12}$ \ergcmsec\ in the 2-10 keV band for the
Seyfert nucleus, corresponding to an intrinsic L$_X$ (2-10 keV) = 7$\times
10^{42}$ \ergsec.  The high energy component from NGC~7319
was already discovered by ASCA (Awaki et al. 1997). The lower spatial
resolution of that observation did not allow the authors to properly
separate the Sey 2 contribution from more extended emission although
all hard emission was assumed to be from the nucleus.  The spectral fit
was similar to the one  derived here except that we find a 
significantly smaller equivalent width.

\section{Unraveling the complex X-ray morphology of SQ}

%------------------------- Figure sketch-------------------
\begin{figure}
\psfig{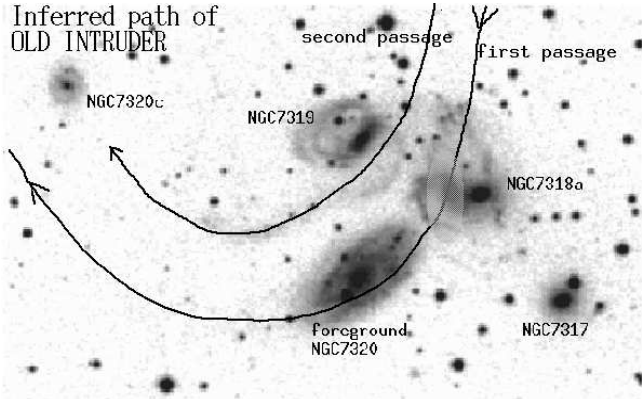}
\caption{A very schematic view of the presumed path of the old intruder
NGC~7320c.  The image of new intruder NGC~7318b is suppressed (shaded out)
since it was not in SQ
at the time.
}
\label{sketch}
\end{figure}
%------------------------- Figure sketch-------------------

A key element in compact group evolutionary history involves stripping
events.  The distribution of neutral gas in SQ is a spectacular example
of these processes that must involve multiple interactions with group
members and/or intruders acquired from the surroundings.  Studies of
the events and conditions in SQ can therefore give insights into
processes that are common in such galaxy aggregates.  

Moles et al.  (1997) proposed a dynamical model for SQ to explain the
observational evidence. Using a larger body of multiwavelength data,
S01 recently evaluated and tested their  ``two intruder'' scenario, that
interprets twin tidal tails as the products of two subsequent passages
of NGC~7320c, the $old$ intruder, through the group (cf.
Fig.~\ref{sketch}).  The $new$ intruder, NGC~7318b, is now entering SQ
with a relative line of sight velocity of $\sim 1000$ km s$^{-1}$.
Some aspects of the two intruder scenario are not shared by all
investigators (e.g. Williams et al. 2002), but there is a consensus
about the evidence of past and present episodes of interactions in SQ, 
that allow us to interpret much of the X-ray
evidence in a self-consistent way. 

The increased sensitivity of Chandra allows us to identify several
components of extended X-ray emission. 1) The NS feature, a clumpy
structure elongated in the NS direction (with possible branches towards
NGC~7319 and the NW) closely coincident with the strongest radio
continuum and optical line emission.  2) An irregular  low surface
brightness  component surrounding the NS feature (radius=$1\farcm5$).
3) Smaller scale extended emission coincident with NGC~7319 and
NGC~7318a. Component 1) is almost certainly shock related while
component 2) could involve shock and/or underlying diffuse emission.
The most straightforward interpretation for components 3) involves an
association with the respective galaxies.  We cannot rule out  a
connection with the recent/ongoing collision because both galaxies,
especially NGC~7318a, may lie in the path of the new intruder.
Individual galaxies are discussed in \S~\ref{galaxies}.

%------------------------- Figure mapcomp-------------------
\begin{figure}
\psfig{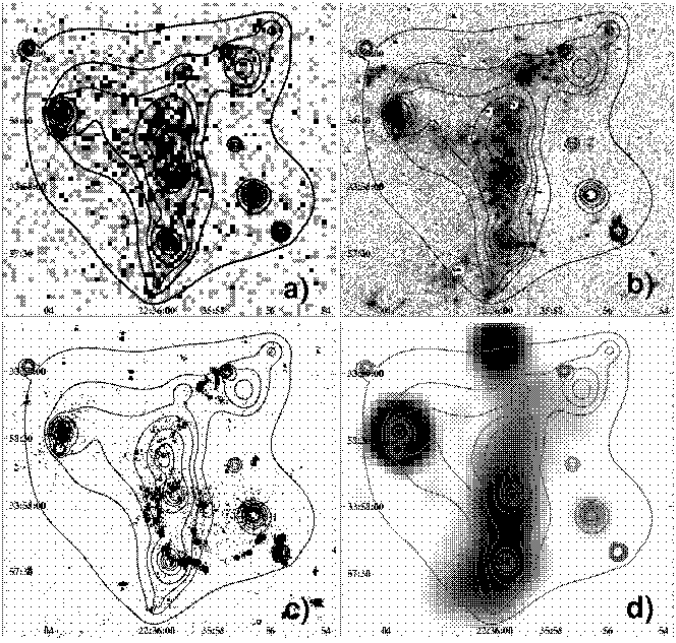}
\caption{X-ray contours (from the 0.5-2.0 keV band adaptively smoothed map)
superimposed on optical emission line and radio continuum data.
Contours are overlayed on: a) X-ray image, binned to $2'' \times 2''$
pixels; b) 
H$\alpha$ emission in the 6500 km s$^{-1}$ SQ velocity range, c) H$\alpha$
emission in the 5700 km s$^{-1}$
new intruder velocity range; d) 1400 MHz radio
continuum emission.  Images have  all the same scale.
}
\label{mapcomp}
\end{figure}
%------------------------- Figure mapcomp-------------------

\subsection{The Shock Zone in SQ}

The NS feature was previously interpreted as evidence for a large scale
shock (Pietsch et al. 1997; S01).  Chandra  data reinforce this view
with better evidence for (see Fig.~\ref{mapcomp}): 
a) spatial coincidence between the NS feature
and similar radio continuum, H$\alpha$ and [\nii] emission structures;
b) lack of coincidence between the X-ray emission and the galaxies or
extended stellar halo in SQ (Moles et al. 1998) and c) lack of X-ray
coincidence with HI line emission in the velocity range of the new
intruder NGC~7318b (S01). 
Optical spectroscopy (Xu et al. 2001) also shows that emission line
ratios from gas coincident with the NS region imply shock conditions.
In the simplest scenario, the shock results from the collision of
NGC~7318b with previously stripped gas in SQ.

In the strong shock approximation, as expected from an 
impact with $v\sim$ 1000 km s$^{-1}$, the shock temperature
\begin{eqnarray}
T_{sh} &=& {3\over 16} (\mu \bar m/k_B) \Delta v^2 \sin^2 \phi \nonumber \\
&\approx& 2.7 \times 10^7 \mu v_8^2 \sin^2 \phi \  \ \ \  {\rm K}
\label{shock-temp}
\end{eqnarray}
The quantities  $\mu$, $\bar m$, $k_B$, and $v_8$ are the mean
molecular weight, mean particle mass of the gas, Boltzmann's constant,
upstream velocity in units of $10^8$ cm s$^{-1}$, respectively.  In a
perpendicular shock, the  angle between the incoming flow direction and
the shock surface $\phi$ = 90$\degr$.  We further assume $\mu$=1,
which accounts for a completely neutral upstream gas and a perfect gas
with ratio of specific heats, $\gamma_c = 5/3$.

This shock  temperature is significantly higher than that derived in
Section~\ref{energydist} and cannot be interpreted as the result of
significant cooling if the collision is ongoing (or even recent, see
later).  It
is more likely that the shock conditions are not as simple as assumed
in Eq.~\ref{shock-temp}, and  require either an oblique (small $\phi$,
cf.  App.~\ref{app:a1}) and/or weak shock (i.e. the upstream medium is
hot and has a sizeable counter-pressure) in which case the strong shock
assumption breaks down.  The effects of a magnetic field are also
considered in App.~\ref{app:a2}, but the field strength derived appears
to be rather high.

\begin{figure}
\psfig{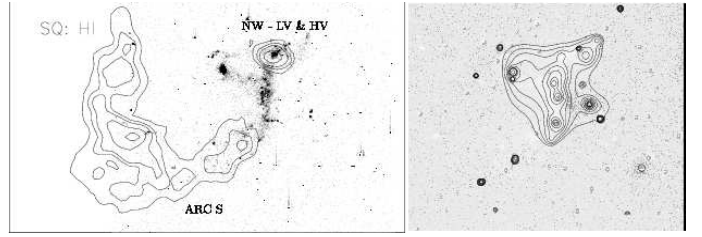}
\caption{Comparison of the spatial distribution of the  neutral hydrogen (LEFT; 
from
Fig. 5 in S01) and X rays (RIGHT) relative to H$\alpha$ in the SQ velocity range
. The X-ray panel covers an area of $\sim 5'\times 4'$}
\label{Xhia}
\end{figure}

If the new intruder collides with a neutral hydrogen cloud, as
suggested by the spatial ``continuity" between the NS feature and
\hi\ clouds N and S of it with consistent velocities (see
Fig.~\ref{Xhia}), we expect strong shock conditions to prevail.  To
reconcile expected and  observed post-shock temperatures,  the incoming
flow should cross the shock at an angle of $\phi \approx 30^\circ$, for
an upstream \hi\ temperature of 100 K (see~App.~\ref{app:a2},
Eq.~\ref{strong_shock-temp_rat}).  Since the upstream Mach number,
$M_\infty \sim 930$, is very high, the opening angle of the Mach cone,
and thus also of the bow shock, should indeed be fairly small.
Therefore we do not expect a significant amount of hotter gas to exist
(e.g.\ in the stagnation point region, where the temperature is at most
a factor of 4.5 higher). Thus the shock conditions look fairly
reasonable, and although they may look somewhat specific, they are
easily realized.

Due to the high Mach number, the compression ratio should be close to
that for a perpendicular shock, e.g.  $n_X/n_1 \approx 4$
(Eq.~\ref{shock-comp_rat}).  To evaluate the gas densities $n_X$, we
can assume the simplified spectral models derived in
Section~\ref{NS-feature}, that will give a good estimate of these
quantities.  
In the NS feature, with a luminosity L$_X \sim 1.5\times
10^{41}$  \ergsec\ in an ellipsoidal volume with $V= {4 \over 3} \pi a
\,b\, c \approx  {4 \over 3} \pi 11'' \times 38'' \times 11''$ ($a, b,
c$ denote the semi-major axes), the density:  \begin{equation} \rm
n_X=\left({L_X\over V_X\Lambda(T_X)} \right)^{1/2} \sim 2.7\times
10^{-2} \ \ \ \  {\rm cm}^{-3} \,, \end{equation} for $\Lambda(T_X)
\approx 7\times 10^{-24}$, appropriate for gas with 10\% solar
metallicity (e.g.\ B\"ohringer \& Hensler 1989).  This suggests a gas
mass M$\rm _{gas} \sim 6.5 \times 10^8$ M$_{\sun}$, and a cooling time
\begin{eqnarray}
t_{\rm cool}  \simeq  {3 k_B T_X \over n_X \Lambda(T_X)} 
\approx 4.2 \times 10^8 \, {\rm yr} 
\label{cooltime}
\end{eqnarray}
The actual density is likely to be locally higher (factors of $\sim
2$), since the emission is obviously clumpy and more compressed towards
the W. In the low surface brightness region surrounding it, the average
density is a factor $\sim  10$ lower, and at roughly the same
temperature as the NS-feature, giving a cooling time also $\sim $ 10
times longer. Thus the shock temperature could be explained  by
cooling only if the shock is $>>$10$^8$ year old 
which is inconsistent with other (all shorter)
timescales (S01).

%----------------------- figure -------------------------
%----------------------- figure -------------------------

We can infer the pre-shock density $n_{\infty}$ $\sim 6.5 \times
10^{-3} \, {\rm cm}^{-3}$ as a mean value of n$_1 \sim 3.4 \times
10^{-3}-1\times 10^{-2}$ cm$^{-3}$ derived for the \hi\  densities in
Arc-S and NW-HV clouds (from the data of Williams et al. 2002).  We therefore
derive a compression ratio  $n_X/n_1 \approx 4$, consistent to the
expected value (Eq.~\ref{shock-comp_rat}).

\subsection{Constraints from UV, optical and radio continuum emission}

The large multiwavelength database available for SQ motivates additional 
considerations:

\noindent $\bullet$ The gas involved in the NS
X-ray/H$\alpha$+[NII]6583/radio continuum feature, and at least some of
the \hi\ clouds, is a product of $past$ $tidal$ interactions. Ram
pressure stripping is not very efficient, given the low densities
involved and the low velocity dispersion in SQ. This suggest that
NGC~7318b can retain a significant residual ISM after the encounter
with SQ:  $n_{IGM} v_{N7381b}^2 \geq 2 n_{N7381b} \sigma_{N7381b}^2$
(Takeda et al. 1984) gives $\sigma_{N7381b} \leq 100$ km s$^{-1}$,
using $n_{IGM} \sim 10^{-2} \, {\rm cm}^{-3}$ for the IGM number
density and $n_{N7381b} \sim 1 \, {\rm cm}^{-3}$ for the average
galaxy's ISM density.  The derived velocity dispersion
$\sigma_{N7381b}$ corresponds to a galaxy mass of less than $\sim 3
\times 10^{10} \, \msol$.

\noindent $\bullet$ The NS feature is spatially (anti)correlated with
the residual \hi\ gas N and S of it. A lot of this gas may be related to
the last passage of NGC~7320c through SQ. The central part of it has
recently been shocked by NGC~7318b.  The transverse velocity component
is unknown and the direction of motion of NGC~7318b is therefore
uncertain.  Striations in the HST and CFHT images (S01) suggest motion
from SW to NE or vice versa. Contour lines of the X-ray emission in NS
feature suggest a higher compression on the W side, favouring motion
from NE to SW.

\noindent $\bullet$ The low surface brightness emission surrounding the
sharper NS feature could be interpreted as a preexisting hot IGM.  
The low velocity dispersion in SQ members requires it
to have been heated by previous collisions (e.g. NGC~7320c?). The 
inferred trajectory in Fig.~\ref{sketch} and a projected distance of 
$\sim 120$ kpc, allow us to estimate the projected velocity $\langle v_{OI} 
\rangle \sim {120\, {\rm kpc} \over 3 \times 10^8 \, {\rm yr}} \approx 
400$ km s$^{-1}$, which is a lower limit. According to
Eq.~(\ref{shock-temp}) this velocity corresponds to $ T \sim 6 \times
10^6$ K, in excellent agreement with the measured X-ray temperature.
The long cooling time derived  above relative to the inferred  last
passage of NGC~7320c    suggests that such a hot IGM component
would not have cooled significantly over this period.  The rather
uniform distribution of the low surface brightness emission (in
contrast to e.g.  the NS feature) would be  due to the equilibration by
shock and/or sound waves that were able to propagate through the IGM
within at least the last $3 \times 10^8 \,{\rm yr}$ (sound crossing
time scale). We expect that the impact of NGC~7318b will produce only a
weak shock here, given  the relatively high gas temperature.  If this
component is related to the ISM of NGC~7318b, rather than residual hot
gas in SQ, we must assume that it has expanded and cooled from an
expected post-shock temperature given by Eq.~(\ref{shock-temp}).

\noindent $\bullet$ The X-ray/radio continuum/optical emission line extension 
or tail towards the NW is a puzzle if the collision is ongoing.
The X-ray cooling timescale is about an order of magnitude longer
than the new intruder crossing  (t$_c$(NGC~7318b)$\sim$ few $\times$10$^7$ 
years). If the former timescale were a better estimate for the 
age of the shock we could more easily account for the NW extension as 
a tidal feature.

\noindent $\bullet$ The X-ray gap shown in Fig.~\ref{cut} is an additional
puzzle although similar features have been observed in richer
environments.  The prominent ``cold front" observed in the cluster
A3776 is interpreted as colder gas moving into a hotter ambient medium
(Vikhlinin et al. 2001). Mazzotta et al.  (2002) identify two
additional features, one of which is an arc-like filamentary X-ray
depression oriented perpendicular to the direction of motion.  Large
scale hydrodynamic instabilities coupled with magnetic field effects
are suggested to explain these morphological peculiarities.  
The surface brightness discontinuity seen in
SQ appears sharper and more prominent than in the other examples.  A
temperature discontinuity is associated with the surface brightness
edges in other examples and we cannot detect such a feature in SQ.  On
the other hand, the temperatures of both the ambient gas and the NS
shock feature in SQ are significantly lower than the gas in A3667. This
could cause a more significant drop in emissivity if the temperature
falls below the energies at which the instruments are sensitive.
Forthcoming observations with XMM-Newton, that will give more spectral
information, will hopefully  provide clues to its nature.

%------------------------- Figure details -------------------
\begin{figure}
\unitlength1.0cm
\begin{picture}(9,3.0)
\thicklines
\put(-0.2,0.5){
\begin{picture}(9,2.5)
\resizebox{9cm}{!}{
}
\end{picture}}
\put(-0.1,0.2){
{ a) HST-b band}
}
\put(2.8,0.2){
{ b) HST-b band}
}
\put(5.7,0.2){
{ c) [\sii]$\lambda$6731+cont}
}
\end{picture}
\caption{Close-up of several interesting regions in SQ. The X-ray contours
are superimposed on an HST WFPC2 B-band image 
in panel a) and b) and a [\sii]$\lambda$6731 + continuum image showing 
NGC~7320 in panel c).
The scale of 15$''$ is given in each panel. 
Source numbers from Table~\ref{sourcelist}.
}
\label{details}
\end{figure}
%------------------------- Figure details -------------------

\section{Discrete Sources in the SQ galaxies}
\label{galaxies}

A few discrete sources can 
be identified with galaxies in the group.

\noindent \underline{NGC~7319}:
NGC~7319 is dominated by a strong heavily absorbed nuclear point source
embedded in fainter extended emission that extends to $\sim 10''$
radius.  Three  additional sources (\# 9, 10, 13) are associated with
this galaxy: source \#9 is near an optical point source and the other
two are projected on the spiral arms but do not appear to coincide with
individual sources or with specific condensations (see
Fig.~\ref{details}).

A connection between interactions and AGN activity has been difficult
to prove conclusively although recent evidence has been advanced for an
excess of Seyfert 2 nuclei in a reasonably complete sample of compact
groups (Coziol et al. 2000).  NGC~7319 shows both evidence for a past
interaction event that stripped it of its ISM (tied to
the last passage of NGC~7320c and the generation of the younger
optical tidal tail)  and near nuclear activity in the form of
an emission line jet (Aoki et al.  1996) and triple lobed radio (Aoki
et al 1999) structure on a 10$''$ scale that cannot be so easily
related to past interaction episodes. One can argue that the
quasi-continuous nature of the tidal perturbations in compact groups
might more efficiently channel gas into nuclei giving rise to phenomena
of this kind.  Interpretation of the extended X-ray emission around
NGC~7319 is complicated by the presence of this optical jet and also by
the nearby unresolved X-ray source (source \#9). The most
straightforward assumption is that source \#9 is unrelated to the
nuclear activity. That assumption may be challenged by the discovery of
other AGN with nearby unresolved X-ray sources (Mrk~3, Morse et al.
1995; NGC~4258, Wilson et al. 2001a): in NGC~4151 (Yang et al, 2001)
and Pictor A (Wilson et al.  2001b).  Source \# 9 is $\sim 8''$ S from
the nucleus (Fig.~\ref{details}) and may coincide with a compact
optical object (\# 54 in  Gallagher et al. 2001).  Its BVI colors are
consistent with a late F to early G main sequence star but the observed
magnitude would put it at $\sim$30 kpc without taking into account any
possible extinction.  It is therefore unlikely to be a  galactic star.
It is embedded in a region where CO is detected (Yun et al. 1997) which
indicates that star formation may be occurring there.  

Given the high X-ray luminosity of X-ray point sources assumed to
belong to NGC~7319, L$_X \ge 5 \times 10^{39}$ \ergsec, they  may be
analogs of the Ultra Luminous X-ray sources (ULX), that are found 
in actively star forming galaxies (Zezas et al. 1999; Roberts \&
Warwick 2000; Fabbiano et al. 2001) and some in more normal spiral galaxies
(Tennant et al.\ 2001, Prestwich 2002).  A
definite interpretation about the origin of ULX is not yet 
available.  Both spectral characteristics and variability
arguments suggest that they are binary accretion sources, although the 
requirement of a sizeable central collapsed object might favour beaming
or anisotropy in the accretion process (King et al.  2001; K\"ording,
Falcke \& Markoff 2002).
Sources 10 and 13 are in a
region of NGC~7319 with colors (S01) consistent with a star formation
event that ended about the time it was stripped of its ISM (a few
times 10$^8$ years ago). 

\noindent \underline{NGC~7318a}. A relatively hard source is coincident
with the nucleus, but a broad band (0.5-5.0 keV) radial profile is
inconsistent with the Chandra Point Spread Function, suggesting that
the emission is extended and likely related to the entire galaxy. This
elliptical galaxy appears to lie directly in the path of the ongoing
collision. It is the only galaxy in SQ other than NGC~7319 that shows
radio emission.  All \hi\ and \hii\ that are found N, W and E of this
galaxy belong to new intruder NGC~7318b.   One can  therefore infer an
``inverse-Cartwheel'' scenario where the disk of NGC~7318b has passed
directly through this galaxy.  While it is tempting to say ``yes'' to
such an hypothesis, two arguments warrant caution:  1) the X-ray (and
radio) properties of NGC~7318a are not unlike more isolated galaxies of
the same type and 2) the high velocity of the new intruder leaves
little time for the formation of interaction induced structure or
activity, unless the collision is ``old", as
discussed earlier. Of course {\it in situ} shocking of gas within the
galaxy is not subject to the latter objection.

\noindent \underline{NGC~7317, NGC~7320c, NGC~7318b}.  None of these
galaxies were formally detected with Chandra.  An enhancement is
visible at the position of NGC~7317. We measure 11$\pm$4 counts in the
0.3-10 keV band within an area with $\sim$4$''$ radius, which 
corresponds to an unabsorbed flux f$_X$ $\sim 5 \times 10^{-15}$
\ergcmsec.  This value  can be used as an upper limit for X-ray emission
from the other two galaxies. This is particularly true for NGC~7318b
where complex emission from the NS feature precludes a direct measure.

\noindent \underline{Isolated bright sources}. X-ray sources \#2, 5,
and possibly 4, are detected close to and possibly associated with
bright emission regions SW and N of NGC~7318ab. All such emission
regions show velocities consistent with the ``b" galaxy (S01).
Source \#5 could be associated with ``putative star" \#127 in
Gallagher et al.  (2001).  An enhancement is also seen coincident with
ISO detected starburst A (Xu et al. 1999) this time with an  SQ
velocity.  The inferred luminosities are relatively high and consistent
with  ULX sources in starforming  regions. 

\noindent \underline{NGC~7320} We find two sources (\#6 and 7) in the
nuclear region of NGC~7320 with luminosities $\sim$ 6 and $\sim 10
\times 10^{37}$ \ergsec.  Source \#6 appears to coincide with the
nucleus (see Fig.~\ref{details}) and would represent a relatively low
luminosity nuclear source comparable to a bright binary system.  A
third source, \#11 at the SE edge of NGC~7320 shows a luminosity of
$\sim2$$\times 10^{38}$ \ergsec.  Both extranuclear sources are most
likely bright accreting binary systems similar to the ones found in
other spiral galaxies.  A more speculative interpretation would relate
source 11 to the old tidal tail passing behind NGC~7320, in which case 
L$_X$ would exceed $\sim$ 10$^{39}$ \ergsec\ at the distance of SQ.

\section{Summary and conclusions}

New high sensitivity and high resolution Chandra
observations confirm the complexity
of the X-ray emission in SQ. The most prominent sources are associated
with a large scale shock that is strongest at low energies (E$<$1.5
keV) and a Seyfert 2 nucleus in NGC~7319 that dominates the emission
above 2 keV.  Additional sources are likely associated with members of
SQ and with the foreground galaxy NGC~7320. Low surface brightness
diffuse emission is also detected  in the core of the system associated
either with a large scale IGM or with the shocked ISM of the new
intruder NGC~7318b.

The complex dynamical history of SQ offers the most plausible
explanation for the large scale NS shock as a collision between a high
velocity intruder and a gaseous debris field produced by earlier
interacting events.  Analytical evaluations of the shock conditions, taking
into account the X-ray morphology and spectrum,  require an oblique shock
propagating into a pre-existing \hi\ cloud.  The alternative is to
postulate that the collision is not ongoing and that the shock has
cooled considerably.

Detailed X-ray analysis of compact groups can also provide us with
further insights into the problem of IGM metal enrichment and, in
particular, whether galaxy interactions rather than galactic winds are
the primary process for entropy and heavy element input into the IGM.
The spectral fits presented here are carried out under the assumption
of collisional ionization equilibrium. They suggest a low IGM
metallicity and a preponderance of tidal interactions over galactic
outflows.  More detailed investigations are not warranted by the
statistics of the Chandra data. Analysis of new XMM-Newton data will
provide more stringent constraints on spectral and dynamical properties
of the collision scenario discussed here.  Stephan's Quintet continues
to be an excellent laboratory for studying dynamics and evolution in
compact groups and represents one of the most useful local analogs of
phenomena  thought to be much more common at high redshift.

\begin{acknowledgements}
GT thanks G. Hasinger, J. Tr\"umper and all colleagues at the
Max-Planck-Institut f\"ur extraterrestrische Physik (MPE) for fruitful
discussions and hospitality while part of this work was done.  GT
acknowledges support from  grants from the Italian Space Agency (ASI).
DB thanks G. Hasinger and the MPE for financial support. JS
acknowledges financial support under NASA grant GO0-1142X.  \\ The
compressed files of the ``Palomar Observatory - Space Telescope Science
Institute Digital Sky Survey" of the northern sky, based on scans of
the Second Palomar Sky Survey are copyright (c) 1993-2000 by the
California Institute of Technology and are distributed by agreement.
All Rights Reserved.

\end{acknowledgements}

\appendix
\section{Shock conditions for interaction of New Intruder with IGM}
\label{app:a}

General shock physics tells
us that a supersonically moving body will have a bow shock at its
leading edge. In the rest frame of the high velocity 
intruder the incoming flow
will enter the shock wave at some angle $\phi$, allowing the downstream
material to be deflected and flow subsonically around the galaxy (e.g.
the NW star forming region).

It is known from aerodynamics that the shape of the bow shock and its
stand-off distance (gap between the nose of the obstacle and the bow
shock) cannot be determined from hydrodynamics alone. It also depends
on the geometry since the fluid equations do not exhibit any
characteristic length scale (for a perfect gas). The bow shock can be
viewed as a transition from a planar/perpendicular (near the stagnation
point region) to  an oblique shock. In the following, we will
quantitatively analyze these two situations. In the planar case, in
which the maximum compression occurs, a magnetic field, parallel to the
shock surface for the maximum effect, will also be considered. In the
next subsections we discuss simple analytic expressions that we
have used for the most plausible scenarios.  We will use the subscripts
``sh'' and ``$\infty$'' for downstream and upstream quantities,
respectively, and an adiabatic gas with ratio of specific heats
$\gamma_c = 5/3$.  $M$ denotes the Mach number, $c$ the speed of sound, and 
$v_{A,\infty}  = \sqrt{B/(4 \pi \rho)}$ the Alfv\'en velocity.

\subsection{Oblique shock}
\label{app:a1}
The compression and the temperature ratio for an oblique shock
associated with a bow shock will be
\begin{eqnarray}
\label{shock-comp_rat}
{\rho_\phi \over \rho_{\infty}} &=& {(\gamma +1) M_{\infty}^2 \sin^2\phi \over
(\gamma - 1) M_{\infty}^2 \sin^2\phi + 2}  \approx  4 \,,
\\
\label{shock-temp_rat}
{T_\phi \over T_{\infty}} &=& {[(\gamma -1) M_{\infty}^2 \sin^2\phi +2]
\over (\gamma + 1)^2 M_{\infty}^2 \sin^2\phi } \quad \times \nonumber \\
& & \times \quad {[2 \gamma M_{\infty}^2 \sin^2\phi - (\gamma -1)]
\over (\gamma + 1)^2 M_{\infty}^2 \sin^2\phi }
\\
\label{strong_shock-temp_rat}
& \approx & {2 \gamma (\gamma -1) \over (\gamma + 1)^2} M_{\infty}^2 \sin^2\phi
\\
\label{shock-temp2}
& = & {5\over 16} M_{\infty}^2 \sin^2\phi \nonumber \,.
\end{eqnarray}
Note that Eq. (\ref{shock-temp_rat})
is the generalization of Eq. (\ref{shock-temp}) for arbitrary
shock strengths.
A detailed numerical calculation, which is
beyond the scope of this paper, will be necessary for a more
quantitative comparison, and it should
include a temperature
structure downstream of the bow shock.

\subsection{Perpendicular MHD shock}
\label{app:a2}
Using the above notation, we can write the compression ratio $r$
and the ratio of gas to magnetic pressure in the upstream plasma
$\beta_{\infty}$  as
\begin{equation}
r = {\rho_{sh} \over \rho_{\infty}} = {v_{\infty} \over v_{sh}} =
{B_{sh} \over B_{\infty}} \,,
\label{comprat1}
\end{equation}
\begin{equation}
\beta_{\infty} = {2 c_{\infty}^2 \over \gamma v_{A,\infty}^2} = {8
\pi \rho_{\infty} k_B T_{\infty} \over B_{\infty}^2 \bar m} \,,
\label{beta1}
\end{equation}
and from the Rankine-Hugoniot conditions we can derive the following
equations:
\begin{eqnarray}
&& 2 (2-\gamma) r^2 - \gamma (1+\gamma) \beta_{\infty} M_{\infty}^2 \nonumber \\
&+& \gamma r
\left\{2+ \beta_{\infty} \left[2 + (\gamma -1) M_{\infty}^2 \right]\right\} = 0
\,,
\label{r-h1}
\end{eqnarray}
and
\begin{equation}
{T_{sh} \over T_{\infty}} = {1\over r}\left[ 1+{1-r^2 \over \beta_{\infty}} +
\gamma \left(1- {1\over r}\right) M_{\infty}^2 \right] \,;
\label{r-h2}
\end{equation}
for $\beta_{\infty} \to \infty$ we recover the ordinary gas shock
relations (see Landau \& Lifshitz 1959). Eq.~(\ref{r-h2}) can be
used to derive the magnetic field strength in the uncompressed gas
due to
\begin{equation}\label{bfield}
    B_{\infty}= \sqrt{{8\pi k_B n_{\infty} T_{\infty} \over (1-r^2)}
    \left[r {T_{sh} \over T_{\infty}} -1
    - \gamma M_{\infty}^2 \left(1-{1\over r} \right)\right]} \,.
\end{equation}
For the shock propagating into the NS feature to be compressive it has
to be a fast MHD shock, satisfying the condition: $v_{N~7318b} \geq
c_{\infty}^2 + v_{A,\infty}^2$. 

If the pre-shock material is the ``missing"
part of an \hi\ structure connecting the Arc-S and NW-HV regions in
Williams et al (2002, see Fig.\ref{Xhia}), its inferred density 
$n_{\infty}$ $\sim 6.5 \times 10^{-3} \, {\rm cm}^{-3}$ 
and upstream temperature
$T_{\infty} = 100$ K.
Thus $c_{\infty} \approx 1.1$
km s$^{-1}$ and the shock is indeed hypersonic since the Mach number is
$M_{\infty} \approx 932$. In case of a strong shock ($M_{\infty}
\gg 1$) Eq.~(\ref{r-h1}) yields a compression ratio of $r=4$ for the
derived X-ray quantities. 
This results
in an unshocked magnetic field strength of $B_{\infty}
\approx 1.13 \times 10^{-5}$ G and thus, in the strong shock
approximation, of a factor $4$ higher in the shocked NS feature. As
the Alfv\'en speed equals $v_{A,\infty}  
\approx 280$ km s$^{-1}$, the fast
MHD shock condition is easily fulfilled. Since $B_{\infty} \propto
1/r$ the magnetic field strength decreases with the compression
ratio $r$; therefore very low compression ratios are disfavored
in a perpendicular MHD shock case. Even the field value for $r=4$
seems to be on the high side.

\end{document}